\documentclass[12pt]{article}
\usepackage{amssymb,amsmath,epsfig,graphicx}
\begin{document}
\title{\bf Noether Symmetry Analysis of Anisotropic Universe in Modified Gravity}

\author{M. Farasat Shamir\thanks{farasat.shamir@nu.edu.pk} and
Fiza Kanwal \thanks{fizakanwal{\_}201@yahoo.com}\\\\
Department of Sciences and Humanities, \\National University of
Computer and Emerging Sciences,\\ Lahore Campus, Pakistan.}

\date{}

\maketitle
\begin{abstract}
In this paper, we study anisotropic universe using Noether symmetries in modified gravity. In particular,
we choose locally rotationally symmetric Bianchi type-$I$ universe for the analysis in $f(R,\mathcal{G})$ gravity, where
$R$ is the Ricci scalar and $\mathcal{G}$ is the Gauss-Bonnet invariant.
Firstly, a model $f(R,\mathcal{G})=f_0R^l+f_1\mathcal{G}^n$ is proposed and the corresponding Noether symmetries are investigated.
Further, we have also recovered the Noether symmetries for $f(R)$ and $f(\mathcal{G})$ theories of gravity.
Secondly, some important cosmological solutions are reconstructed. Exponential and power law solutions are reported for a well-known $f(R,\mathcal{G})$ model, 
i.e., $f(R,\mathcal{G})=f_0R^n\mathcal{G}^{1-n}$. Especially, the Kasner's solution is recovered and it is anticipated
that the familiar de-Sitter spacetime giving $\Lambda CDM$ cosmology may be reconstructed for some suitable value of $n$.
\end{abstract}

{\bf Keywords:} Anisotropic Models; Exact Solutions; Noether Symmetry.\\
{\bf PACS:} : 04.20.Jb; 98.80.-k; 98.80.Jk.

\section{Introduction}

Experimental observations in the recent years have indicated that
universe is expanding \cite{1}-\cite{4}. The candidate referred for
the explanation of this expansion is known as Dark energy (DE). DE
is thought to be the energy density reserved in the cosmological
constant and its value is considered $\rho_{\Lambda}\sim 10^{-3}
eV^4$ \cite{cop}. Modification of general relativity (GR) can also
be one of the possibility to explain the existence of DE. The
presence of some additional terms in the matter part of gravitational
action which cause a minimum modification in GR can be helpful to
explore the nature of DE \cite{cog}. Another method to explain the
cosmological acceleration is to modify the geometrical part of the
action, by coupling curvature scalars, topological invariants and
their derivatives which results in modified theories like $f(R)$
gravity, $f(\mathcal{G})$ gravity, $f(R,T)$ gravity etc., where $R$
is the Ricci scalar and $\mathcal{G}$ is the Gauss-Bonnet invariant,
and $T$ is the trace of energy momentum tensor.

A new modified theory by considering a general
function of $R$ and $\mathcal{G}$ as $f(R,\mathcal{G})$ has gained popularity in recent years
\cite{noj}-\cite{noj2}.
There are many interesting aspects of Gauss-Bonnet theory which motivate the researchers to study
modified theories of gravity involving Gauss-Bonnet term. In particular, it have been
shown that Gauss-Bonnet gravity can address the DE problem
without the need for any exotic matter components \cite{noj3}.
Gauss-Bonnet term is a specific combination of
curvature invariants that includes Ricci scalar, Ricci and Riemann tensors.
In fact, Gauss-Bonnet invariant naturally arises in the process of
quantum field theory regularization and renormalization of
curved spacetime. In particular, including $\mathcal{G}$ and $R$ in a bivariate function provides
a double inflationary scenario where the two acceleration phases are led by $\mathcal{G}$ and $R$ respectively \cite{mar}.
Moreover, the involvement of Gauss-Bonnet invariant may play an important role in the early time
expansion of universe as it is connected with the string theory and the trace anomaly \cite{cap.lau}.
The viability of
modified Gauss-Bonnet gravity has been studied by considering
different realistic models using the weak energy
condition and it was concluded that $f(R,\mathcal{G})$ gravity
models show consistency with the recent observational data \cite{ata}.

Symmetry approach performs a pivotal part to find exact solutions or
simply reduce a non-linear system of equations to a linear system of
equations. In the literature (see \cite{wei}-\cite{cam1} for
references), Noether symmetries have been studied in the context of
cosmology and astrophysics, in particular,  to investigate the exact
solutions of field equations. Noether symmetry is a  proficient
method to calculate unknown variables of differential equations.
Sharif and Waheed \cite{sha1} studied Bardeen model and stringy
charged black holes by using approximate symmetry methods. They also
explored Noether symmetries of Friedmann-Robertson-Walker (FRW) and
locally rotationally symmetric (LRS) Bianchi type-$I$ (BI) universe
models by adding an inverse curvature term in Brans-Dicke theory
\cite{sha2}. Capozziello et al. \cite{cap}
discussed FRW universe model in $f(R,\mathcal{G})$ gravity using the
Noether symmetry approach. In modified scalar-tensor gravity, Sharif
and Shafique \cite{sha3} studied BI model using Noether and Noether
gauge symmetry. Sharif and Fatima \cite{sha4} investigated Noether
symmetry of flat FRW model for vacuum and non-vacuum cases in $f(G)$
gravity.

The spatially homogeneous but largely anisotropic nature of early
universe was disclosed after the discovery of cosmic microwave
background radiation (CMBR) \cite{ell}. Bianchi type universe models
can be considered to quantify the change of anisotropy in the early
universe through recent observations. These universe model indicates
that the anisotropy of early universe determines the
acceleration rate of the universe. If the primary anisotropy is less than
this rate of acceleration would lead to a highly isotropic universe
\cite{bar}. Akarsu and Kilinc \cite{aka} considered BI model to study
different equation of state (EoS) models which coincide with
de-Sitter universe. Sharif and Zubair \cite{sha} studied the solutions
of BI universe model by using power-law and exponential expansions
in scalar-tensor gravity. Shamir \cite{shm} explored exact solutions
of LRS BI universe model and investigated
physical behavior of cosmological parameters in $f(R,T)$ gravity.
Shamir and Ahmad \cite{shm1} has discussed the Noether symmetry
approach for FRW universe model in $f(\mathcal{G},\mathcal{T})$
gravity. In another paper \cite{shm6}, the same authors explored some important cosmological
solutions in $f(\mathcal{G}, \mathcal{T})$ gravity using Noether
symmetries. In particular they found some interesting results by considering LRS BI spacetime and 
recovered $\Lambda CDM$ model universe for some specific choice of $f(\mathcal{G}, \mathcal{T})$ gravity model. Thus it seems interesting
to investigate Noether symmetry of anistropic universe in modified
gravity.

In this paper, we have explored Noether symmetries of BI cosmological
model. The interesting physical forms of $f(R,\mathcal{G})$ can be
determined by existence of Noether symmetry which allows to reduce
the dynamics. The Noether symmetry approach has been  broadly used
for modified theories which gave some applicable results for
cosmological systems. In this work, exponential and power law solutions are reported for a well-known $f(R,\mathcal{G})$ model, i.e., $f(R,\mathcal{G})=f_0R^n\mathcal{G}^{1-n}$. Especially, the Kasner's solution is recovered and it is anticipated
that the familiar de-Sitter spacetime giving $\Lambda CDM$ cosmology may be reconstructed for some suitable value of $n$.
The layout of paper is as follows: In section
\textbf{2}, we give gravitational action and Einstein field
equations for modified Gauss-Bonnet gravity and  derive a point like canonical Lagrangian for configuration space.
We explore the Noether symmetries for some cosmological models in section \textbf{3}. In section
\textbf{4}, we have discussed some examples of exact solution in cosmological context. Final
remarks are given in last section.

\section{Modified Field Equations and Lagrangian Framework}
For $f(R,\mathcal{G})$ gravity the most general action in
4-dimensions is \cite{cap}
\begin{equation}\label{1}
\mathcal{A}= \int d^{4}x
\sqrt{-g}\bigg[\frac{1}{2\kappa^{2}}f(R,\mathcal{G})+\mathcal{L}_{m}\bigg],
\end{equation}
where $\kappa$ is the coupling constant,  $\mathcal{L}_{m}$ is the
matter Lagrangian and $g$ denotes the determinant of the metric tensor.
$\mathcal{G}$ indicates the Gauss-Bonnet invariant
\begin{equation}
\mathcal{G}\equiv
R^{2}-4R_{\mu\nu}R^{\mu\nu}+R_{\mu\nu\lambda\sigma}R^{\mu\nu\lambda\sigma},
\end{equation}
where $R_{\mu\nu}$ is Ricci tensor, $R_{\mu\nu\lambda\sigma}$
represents Riemann tensor and $R=g^{\mu\nu}R_{\mu\nu}$ is called the
Ricci scalar. From now onwards, we will use $f(R,\mathcal{G})\equiv
f$, $\frac{\partial f(R,\mathcal{G})}{\partial \mathcal{G}}\equiv
f_\mathcal{G}$, $\frac{\partial f(R,\mathcal{G})}{\partial R}\equiv
f_R$, etc. Variation of (\ref{1}) with respect to the metric tensor
$g_{\mu\nu}$, leads to the modified field equations \cite{lau}
\begin{align}\nonumber
0=&\kappa^{2}T^{\mu\nu}+\frac{1}{2}g^{\mu\nu}f-2f_\mathcal{G}RR^{\mu\nu}+4f_\mathcal{G}R^{\mu}_{\rho}R^{\nu\rho}-2f_\mathcal{G}R^{\mu\rho\sigma\tau}
R^{\nu}_{\rho\sigma\tau}-4f_\mathcal{G}R^{\mu\rho\sigma\nu}R_{\rho\sigma}\\\nonumber&+2R\nabla^{\mu}\nabla^{\nu}f_\mathcal{G}-2g^{\mu\nu}R
\nabla^{2}f_\mathcal{G}-4\nabla_{\rho}\nabla^{\mu}f_\mathcal{G}R^{\nu\rho}-4R^{\mu\rho}\nabla_{\rho}\nabla^{\nu}f_\mathcal{G}
+4R^{\mu\nu}\nabla^{2}f_\mathcal{G}\\&+4g^{\mu\nu}R^{\rho\sigma}\nabla_{\rho}\nabla_{\sigma}f_\mathcal{G}
-4R^{\mu\rho\nu\sigma}\nabla_{\rho}\nabla_{\sigma}f_\mathcal{G}-f_\mathcal{G}R^{\mu\nu}+\nabla^{\mu}\nabla^{\nu}f_\mathcal{G}
-g^{\mu\nu}\nabla^{2}f_\mathcal{G},
\end{align}
where $\nabla$ represents the covariant derivative and the energy
momentum tensor $T_{\mu\nu}$ is defined as
$$T_{\mu\nu}=\frac{-2}{\sqrt{-g}}\frac{\delta(\sqrt{-g}\mathcal{L}_{m})}{\delta g^{\mu\nu}}.$$
Let us note that Einstein equations for GR are recovered by putting
$f(R,\mathcal{G})=R$ and by replacing the $f(R,\mathcal{G})$ with
$f(\mathcal{G})$, we obtain the field equations for $f(\mathcal{G})$ gravity.
We consider an LRS BI universe model, defined by the line element
\cite{shm2}
\begin{equation}
ds^{2}=dt^{2}-A(t)^2dx^{2}-B(t)^{2}(dy^{2}+dz^{2}), \end{equation}
where $A$ and $B$ are known as cosmic scale factors of the universe.
We need to find out a point-like canonical Lagrangian
$\mathcal{L}(q^i,\dot{q}^i)$ from the gravitational
action, defined on the configuration space $Q$ and the corresponding
tangent space $TQ$ where $q^{i}$ represents $n$ generalized
positions and dot denotes time derivative. By using the technique of
the Lagrange multipliers, we can deduce $R$ and $\mathcal{G}$ as
constraints for dynamics. In order to reduce the order of derivatives in Lagrangian $\mathcal{L}$, we use the Lagrange
multiplier approach and integrate by parts so that the Lagrangian
$\mathcal{L}$ becomes canonical \cite{cap6}.
We rewrite the action (\ref{1}) as
\begin{equation}\label{2}
\mathcal{A}= 2\pi^{2}\int
dtAB^{2}\bigg[f(R,\mathcal{G})-\chi_{1}\bigg(R+2\bigg(\frac{\ddot{A}}{A}+2\frac{\ddot{B}}{B}+2\frac{\dot{A}\dot{B}}{AB}+
\frac{\dot{B}^{2}}{B^{2}}\bigg)\bigg)
-\chi_{2}\bigg(\mathcal{G}-8\bigg(\frac{\ddot{A}\dot{B}^{2}}{AB^2}+2\frac{\dot{A}\dot{B}\ddot{B}}{AB^2}\bigg)\bigg)\bigg],
\end{equation}
where dot indicates the derivative with respect to time $t$ and
$\chi_{1},\chi_{2}$ are the Lagrange multipliers. The Ricci scalar and the Gauss-Bonnet invariant for LRS BI
metric is \cite{shm2}
\begin{eqnarray}
R=-2\bigg(\frac{\ddot{A}}{A}+2\frac{\ddot{B}}{B}+2\frac{\dot{A}\dot{B}}{AB}+\frac{\dot{B}^{2}}{B^{2}}\bigg),&&
   \mathcal{G}=8\bigg(\frac{\ddot{A}\dot{B}^{2}}{AB^2}+2\frac{\dot{A}\dot{B}\ddot{B}}{AB^2}\bigg).
\end{eqnarray}
The Lagrange multipliers $\chi_{1},\chi_{2}$ are obtained by varying
the action $(5)$ with respect to $R$ and $\mathcal{G}$ respectively as
\begin{eqnarray}\label{3}
\chi_{1}=f_R,&& \chi_{2}=f_\mathcal{G}.
\end{eqnarray}
Using Eq.(\ref{3}), the action (\ref{2}) becomes
\begin{equation}
\mathcal{A}=2\pi^{2}\int
dtAB^{2}\bigg[f-f_R\bigg(R+2\bigg(\frac{\ddot{A}}{A}+2\frac{\ddot{B}}{B}+2\frac{\dot{A}\dot{B}}{AB}+\frac{\dot{B}^{2}}{B^{2}}\bigg)\bigg)-f_\mathcal{G}
\bigg(\mathcal{G}-8\bigg(\frac{\ddot{A}\dot{B}^{2}}{AB^2}+2\frac{\dot{A}\dot{B}\ddot{B}}{AB^2}\bigg)\bigg)\bigg],
\end{equation}
After an integration by parts, the Lagrangian turns out to be
\begin{equation}
\mathcal{L}=2A\dot{B}^{2}f_R+4\dot{A}\dot{B}Bf_R+2\dot{A}B^{2}\frac{d}{dt}f_R
+4AB\dot{B}\frac{d}{dt}f_R-8\dot{A}\dot{B}^{2}\frac{d}{dt}f_\mathcal{G}+AB^{2}(f-Rf_R-\mathcal{G}f_\mathcal{G}).\label{lag1}
\end{equation}
This is a point-like canonical Lagrangian whose configuration space
is $\mathcal{Q}\equiv\{A,B,R,\mathcal{G}\}$ and tangent space is
$\mathcal{QT}\equiv\{A,B,R,\mathcal{G},\dot{A},\dot{B},\dot{R},\dot{\mathcal{G}}\}$.
Due to highly nonlinear nature of Lagrangian, it is complicated to
deal with it. For the sake of simplicity, we assume $B=A^{m}$.
The physical importance of this assumption is that it gives constant ratio of shear and expansion scalar
\cite{col}. Thus the Lagrangian (\ref{lag1}) takes the form
\begin{eqnarray}\nonumber
\mathcal{L}&=&2m^2A^{2m-1}\dot{A}^{2}f_R+4mA^{2m-1}\dot{A}^{2}f_R+2A^{2m}\dot{A}\frac{d}{dt}f_R+4mA^{2m}\dot{A}\frac{d}{dt}f_R
\\&&-8m^2A^{2m-2}\dot{A}^{3}\frac{d}{dt}f_\mathcal{G}+A^{2m+1}[f-Rf_R-\mathcal{G}f_\mathcal{G}],\label{lag2}
\end{eqnarray}
which is now a function of $A$, $R$ and $\mathcal{G}$.

\section{Noether Symmetry and $f(R,\mathcal{G})$ Gravity}
In the presence of Noether symmetry, the constants of motion can be
selected by the reduction of dynamical system \cite{cap2}. We
consider the vector field and its first prolongation respectively as
\begin{equation}
X=\xi(t,q^j)\frac{\partial}{\partial
t}+\eta^i(t,q^j)\frac{\partial}{\partial q^j},
\end{equation}
\begin{equation}
X^{[1]}=X+(\eta^i_{,t}+\eta^i_{,j}\dot{q}^j-\xi_{,t}
\dot{q}^i-\eta_{,j}\dot{q}^j\dot{q}^i)\frac{\partial}{\partial
\dot{q}^j},
\end{equation}
where $\xi$ and $\eta$ are the coefficients of the generators, $q^i$
provides the $n$ number of positions and dot gives the derivative
with respect to time $t$ \cite{haw}. The vector field $X$ produces
Noether gauge symmetry provided the condition
\begin{equation}
X^{[1]}\mathcal{L}+(D\xi)\mathcal{L}=DG(t,q^i)
\end{equation}
is preserved. Here $G(t,q^i)$ denotes gauge term and $D$ is an
operator defined as
\begin{equation}
D=\frac{\partial}{\partial t}+\dot{q}^i\frac{\partial}{\partial
q^i}.
\end{equation}
The Euler-Lagrange equations are given by \cite{sha5}
\begin{equation}\frac{\partial\mathcal{L}}{\partial
q^{i}}-\frac{d}{dt}\bigg(\frac{\partial\mathcal{L}}{\partial\dot{q}^{i}}\bigg)=0.\label{el}
\end{equation}
Contraction of Eq.(\ref{el}) with some unknown function
$\psi^i\equiv\psi^i(q^j)$ yields
\begin{equation}
\psi^i\bigg(\frac{\partial\mathcal{L}}{\partial
q^{i}}-\frac{d}{dt}\bigg(\frac{\partial\mathcal{L}}{\partial\dot{q}^{i}}\bigg)\bigg)=0,\label{ell}
\end{equation} It is easy to verify that
\begin{equation}
\frac{d}{dt}\bigg(\psi^i\frac{\partial\mathcal{L}}{\partial
\dot{q}^i}\bigg)-\bigg(\frac{d}{dt}\psi^i\bigg)\frac{\partial\mathcal{L}}{\partial
\dot{q}^i}=\psi^i\frac{d}{dt}\bigg(\frac{\partial\mathcal{L}}{\partial
\dot{q}^i}\bigg).
\end{equation}
Putting this value in Eq.(\ref{ell}) provides us with
\begin{equation}
L_{X}\mathcal{L}=\bigg(\frac{d}{dt}\psi^i\bigg)\frac{\partial\mathcal{L}}{\partial
\dot{q}^i}+\psi^i\frac{d}{dt}\bigg(\frac{\partial\mathcal{L}}{\partial
\dot{q}^i}\bigg)=\frac{d}{dt}\bigg(\psi^i\frac{\partial\mathcal{L}}{\partial
\dot{q}^i}\bigg),
\end{equation} where $L_{X}$ is the Lie derivative with respect to the Noether
vector $X$. Lagrangian generates a Noether symmetry if Lie
derivative, for a vector field $X$, vanishes
\begin{eqnarray}
L_{X}\mathcal{L}=0.
\end{eqnarray}
The energy condition is given as
\begin{equation}\label{energycondition}
\sum_i\dot{q}^i
\frac{\partial\mathcal{L}}{\partial\dot{q}^i}=E_\mathcal{L}.
\end{equation}
We derive the Euler-Lagrange equations (\ref{el}) for $A$, $R$,
$\mathcal{G}$ respectively as,
\begin{eqnarray}\nonumber
&&2m(2m^2+3m-2)A^{2m-2}\dot{A}^{2}f_R+4m(m+2)A^{2m-1}\ddot{A}f_R+4m^2A^{2m-1}\dot{A}\frac{d^2}{dt^2}f_R\\\nonumber
&&-32m^2(m-1)A^{2m-3}\dot{A}^{3}\frac{d}{dt}f_\mathcal{G}+2(2m+1)A^{2m}\frac{d^2}{dt^2}f_R-48m^2A^{2m-2}\dot{A}\ddot{A}\frac{d}{dt}f_\mathcal{G}\\\nonumber
&&+4(m+1)A^{2m-1}\dot{A}\frac{d}{dt}f_R-24m^2A^{2m-2}\dot{A}^{2}\frac{d^2}{dt^2}f_\mathcal{G}-(2m+1)A^{2m}\big[f-Rf_R\\&&-\mathcal{G}f_\mathcal{G}\big]=0,\label{el1}
\end{eqnarray}
\begin{eqnarray}\nonumber
&&\bigg[R+2\bigg(\frac{\ddot{A}}{A}+3m^2\bigg(\frac{\dot{A}}{A}\bigg)^{2}+2m\frac{\ddot{A}}{A}\bigg)\bigg]f_{RR}+
\bigg[ \mathcal{G}
+8\bigg(2m^2\bigg(\frac{\dot{A}}{A}\bigg)^{4}-2m^3\bigg(\frac{\dot{A}}{A}\bigg)^{4}
\\&&-3m^2\frac{\dot{A}^{2}\ddot{A}}{A^3}\bigg)\bigg]f_{R\mathcal{G}}=0,\label{el2}
\end{eqnarray}
\begin{eqnarray}\nonumber
&&\bigg[R+2\bigg(\frac{\ddot{A}}{A}+3m^2\bigg(\frac{\dot{A}}{A}\bigg)^{2}+2m\frac{\ddot{A}}{A}\bigg)\bigg]f_{R\mathcal{G}}+
\bigg[ \mathcal{G}
+8\bigg(2m^2\bigg(\frac{\dot{A}}{A}\bigg)^{4}-2m^3\bigg(\frac{\dot{A}}{A}\bigg)^{4}
\\&&-3m^2\frac{\dot{A}^{2}\ddot{A}}{A^3}\bigg)\bigg]f_{\mathcal{G}\mathcal{G}}=0.\label{el3}
\end{eqnarray}
This is important to note that (\ref{el2}) and (\ref{el3}) are
symmetric. The energy condition (\ref{energycondition}) takes the
form
\begin{align}
\bigg(\frac{\dot{A}}{A}\bigg)^{2}f_R+\frac{\dot{A}}{A}\frac{d}{dt}f_R-4\bigg(\frac{\dot{A}}{A}\bigg)^{3}\frac{d}{dt}f_{\mathcal{G}}
-\frac{1}{6}\bigg[f-Rf_R-\mathcal{G}f_{\mathcal{G}}\bigg]=0.
\end{align} The corresponding vector field becomes
\begin{eqnarray}
X=\alpha\frac{\partial}{\partial A}+\beta\frac{\partial}{\partial
R}+\gamma\frac{\partial}{\partial
\mathcal{G}}+\dot{\alpha}\frac{\partial}{\partial
\dot{A}}+\dot{\beta}\frac{\partial}{\partial
\dot{R}}+\dot{\gamma}\frac{\partial}{\partial
\dot{\mathcal{G}}},\label{vectorfeild}
\end{eqnarray}
The symmetry generators $\alpha$, $\beta$ and $\gamma$ are functions of $A$, $R$ and
$\mathcal{G}$.
Using Lagrangian (\ref{lag2}) and Noether equation
(\ref{vectorfeild}), we get a system of partial differential
equations (PDEs)
\begin{eqnarray} \nonumber
&&2m^3\alpha f_R + 3m^2\alpha f_R-2m\alpha f_R+m^2\beta
Af_{RR}+2m\beta Af_{RR}+m^2\gamma
Af_{R\mathcal{G}}\\\nonumber&&+2m\gamma Af_{R\mathcal{G}}+
2m^2A\frac{\partial\alpha}{\partial
A}f_R+4mA\frac{\partial\alpha}{\partial
A}f_R+A^2\frac{\partial\beta}{\partial
A}f_R+2mA^2\frac{\partial\beta}{\partial
A}f_{RR}\\&&+A^2\frac{\partial\gamma}{\partial
A}f_{R\mathcal{G}}+2mA^2\frac{\partial\gamma}{\partial
A}f_{R\mathcal{G}}=0,\label{N1}
\end{eqnarray}
\begin{eqnarray} \nonumber
&&2m\alpha f_{RR}+4m^2\alpha f_{RR}+2m\beta Af_{RRR}+\beta
Af_{RRR}+\gamma Af_{RR\mathcal{G}}+2m\gamma
Af_{RR\mathcal{G}}\\\nonumber &&+A\frac{\partial\alpha}{\partial
A}f_{RR}+2mA\frac{\partial\alpha}{\partial
A}f_{RR}+2m^2A\frac{\partial\alpha}{\partial
R}f_R+4m\frac{\partial\alpha}{\partial
R}f_R+A\frac{\partial\beta}{\partial R}f_{RR}+\\\label{N2}
&&2mA\frac{\partial\beta}{\partial
R}f_{RR}+A\frac{\partial\gamma}{\partial
R}f_{R\mathcal{G}}+2mA\frac{\partial\gamma}{\partial
R}f_{R\mathcal{G}}=0,
\end{eqnarray}
\begin{eqnarray}\nonumber&& 2m\alpha
f_{R\mathcal{G}}+4m^2\alpha f_{R\mathcal{G}}+\beta
Af_{RR\mathcal{G}}+2m\beta Af_{R\mathcal{G}\mathcal{G}}+\gamma
Af_{RR\mathcal{G}}+2m\gamma Af_{R\mathcal{G}\mathcal{G}}\\\nonumber
 &&+A\frac{\partial\alpha}{\partial
A}f_{R\mathcal{G}}+2mA\frac{\partial\alpha}{\partial
A}f_{R\mathcal{G}}+2m^2\frac{\partial\alpha}{\partial\mathcal{G}}f_R+4m\frac{\partial\alpha}{\partial\mathcal{G}}f_{R}
+2mA\frac{\partial\beta}{\partial\mathcal{G}}f_{RR}
\\&&+A\frac{\partial\beta}{\partial\mathcal{G}}f_{RR}+A\frac{\partial\gamma}{\partial\mathcal{G}}f_{R\mathcal{G}}
+2mA\frac{\partial\gamma}{\partial\mathcal{G}}f_{R\mathcal{G}}=0,\label{N3}\end{eqnarray}
\begin{eqnarray}\nonumber
&&2m\alpha f_{R\mathcal{G}}-2\alpha f_{R\mathcal{G}}+\beta
Af_{RR\mathcal{G}}+\gamma
Af_{R\mathcal{G}\mathcal{G}}+3A\frac{\partial\alpha}{\partial
A}f_{R\mathcal{G}}+A\frac{\partial\beta}{\partial
R}f_{R\mathcal{G}}\\&&+A\frac{\partial\gamma}{\partial
R}f_{\mathcal{G}\mathcal{G}}=0,\label{N4}\end{eqnarray}
\begin{eqnarray}\nonumber &&2m\alpha f_{\mathcal{G}\mathcal{G}}-2\alpha
f_{\mathcal{G}\mathcal{G}}+\beta Af_{R\mathcal{G}\mathcal{G}}+\gamma
Af_{\mathcal{G}\mathcal{G}\mathcal{G}}+3A\frac{\partial\alpha}{\partial
A}f_{\mathcal{G}\mathcal{G}}+A\frac{\partial\beta}{\partial\mathcal{G}}f_{R\mathcal{G}}\\&&+A\frac{\partial\gamma}{\partial
\mathcal{G}}f_{\mathcal{G}\mathcal{G}}=0,\label{N5}\end{eqnarray}
\begin{eqnarray}
\frac{\partial\alpha}{\partial
R}f_{R\mathcal{G}}+2m\frac{\partial\alpha}{\partial
R}f_{R\mathcal{G}}+\frac{\partial\alpha}{\partial
\mathcal{G}}f_{RR}+2m\frac{\partial\alpha}{\partial
\mathcal{G}}f_{RR}=0,\label{N6}\end{eqnarray}
\begin{eqnarray}  \frac{\partial\alpha}{\partial
R}f_{R\mathcal{G}}=0,~~~~~~~~\frac{\partial\alpha}{\partial
R}f_{\mathcal{G}\mathcal{G}}+\frac{\partial\alpha}{\partial
\mathcal{G}}f_{R\mathcal{G}}=0, \label{N7}\end{eqnarray}

\begin{eqnarray}\frac{\partial\alpha}{\partial
R}f_{RR}+2m\frac{\partial\alpha}{\partial
R}f_{RR}=0,~~~~~\frac{\partial\alpha}{\partial
\mathcal{G}}f_{R\mathcal{G}}+2m\frac{\partial\alpha}{\partial\mathcal{G}}f_{R\mathcal{G}}=0,\label{N8}\end{eqnarray}
\begin{eqnarray}\frac{\partial\alpha}{\partial
\mathcal{G}}f_{\mathcal{G}\mathcal{G}}=0,~~~~~~
\frac{\partial\beta}{\partial
A}f_{R\mathcal{G}}+\frac{\partial\gamma}{\partial
A}f_{\mathcal{G}\mathcal{G}}=0,\label{N9}\end{eqnarray}
\begin{eqnarray}\nonumber(2m+1)\alpha[f-Rf_R -\mathcal{G}f_{\mathcal{G}}]-\beta A[Rf_{RR} +\mathcal{G}f_{R\mathcal{G}}]
-\gamma
A[Rf_{R\mathcal{G}}+\mathcal{G}f_{\mathcal{G}\mathcal{G}}]=0.
\\\label{N10}\end{eqnarray}
The obtained system of PDEs is
over-determined. Hence one can solve it by assigning the suitable
values to the unknown function $f(R,\mathcal{G})$. Here we propose
$f(R,\mathcal{G})$ as linear combination of power law forms of $R$
and $\mathcal{G}$ as
\begin{equation}
f(R,\mathcal{G})=f_0R^l+f_1\mathcal{G}^n,
\end{equation}
where $f_0,~f_1$ are the
arbitrary constants and $l, n$ are any non-zero real numbers.
We get a number of solutions as follows:
\begin{eqnarray}\label{sol1}
l=1,~~~~n=1,~~\alpha=c_1A^{-m+\frac{1}{2}},~~\beta=0,~~\gamma=0,
\end{eqnarray}
\begin{eqnarray}\nonumber&&
l\neq 1,~~~~~ l=\frac{4m^2+4m+1}{3m^2+2m+1},~~n=1,~~\alpha=c_2A^{-\frac{3m^2}{2m+1}},
\\&&\beta=-\frac{c_2A^{-\frac{3m^2}{2m+1}}R(3m^2+2m+1)}{A(2m+1)},~~\gamma=0.\label{sol4}
\end{eqnarray}
Here we have explored some additional symmetries and it would be worthwhile to mention here that Eq.(\ref{sol1}) agree with \cite{cap} for
special case when $m=1$.
The corresponding Lagrangian for the particular functional form $f(R,\mathcal{G})=f_{0}R+f_{1}\mathcal{G}$
becomes
\begin{equation}\mathcal{L}=2m(m+2)f_0A^{2m-1}\dot{A}^2.\end{equation} The
Euler Lagrange equations and energy equation are calculated as
\begin{equation}
2\frac{\ddot{A}}{A}+(2m-1)\bigg(\frac{\dot{A}}{A}\bigg)^{2}=0,
\end{equation}
\begin{equation}
R+2\bigg((2m+1)\frac{\ddot{A}}{A}+3m^2\frac{\dot{A}^{2}}{A^{2}}\bigg)=0,
\end{equation}
\begin{equation}
\mathcal{G}-8\bigg(m(2m+1)\frac{\ddot{A}\dot{A}^{2}}{A^{3}}+2m^2(m-1)\frac{\dot{A}^{4}}{A^4}\bigg)=0,
\end{equation}
\begin{equation}
\bigg(\frac{\dot{A}}{A}\bigg)^{2}=0.
\end{equation}
In these equations the Gauss-Bonnet term $\mathcal{G}$ disappears so
this theory becomes nothing but the General Relativity. If we
consider vacuum case then we obtain Minkowski spacetime.

\subsection{Recovering Noether Symmetries in $f(R)$ and $f(\mathcal{G})$ Theories of Gravity}

Here we investigate the solution of determining system of equations
mainly for two different cases, for $f(R)$ gravity and
$f(\mathcal{G})$ gravity. First we explore solutions for
$f(R,\mathcal{G})=f_0R^{n}$. The
solutions of determining equations are given as
\begin{eqnarray}\nonumber
m=\pm\frac{n-2+\sqrt{3n-2n^2}}{3n-4},~~~~~~~\alpha=c_3A^{\frac{-5n\pm2\sqrt{3n-2n^2}+3n^2}{3n^2-4n}},
\\\beta=-\frac{c_3A^{\frac{-5n\pm2\sqrt{3n-2n^2}+3n^2}{3n^2-4n}}R(n\pm2\sqrt{3n-2n^2})}{(3n-4)An}.\label{sol13}\end{eqnarray}
It is worthwhile to mention here that for a special case when  $n={\frac{3}{2}}$, we obtain
\begin{equation}m=1,~~\alpha=\frac{c_3}{A},~~\beta=-\frac{2c_3R}{A^2},\label{sol9}\end{equation}
and the results agree with \cite{hus, shm12}. To recover Noether
symmetries in $f(\mathcal{G})$ gravity, we choose
$f(R,\mathcal{G})=f_0\mathcal{G}^n$ and the solution in this case
turn out to be as\begin{equation}m=-\frac{1}{2},~~\alpha=c_4 A,~~~~~
\gamma=0.\label{sol14}\end{equation}

\section{Some Exact Cosmological Solutions}

Here we reconstruct some important cosmological solution using an
interesting $f(R,\mathcal{G})$ model, i.e.,
$f(R,\mathcal{G})=f_0R^n\mathcal{G}^{1-n}$ \cite{cap}. For the
simplest non-trivial case, we choose $n=2$. In this case, the
point-like Lagrangian (\ref{lag2}) takes the form
\begin{eqnarray}\nonumber
\mathcal{L}&=&\frac{4f_{0}\dot{A}}{\mathcal{G}}\bigg[(m^2+2m)A^{2m-1}\dot{A}R+(2m+1)A^{2m}\dot{R}-(2m+1)A^{2m}\dot{\mathcal{G}}\frac{R}{\mathcal{G}}\\&&
+4m^2A^{2m-2}\dot{A}^{2}\dot{R}\frac{R}{\mathcal{G}}-4m^2A^{2m-2}\dot{A}^{2}\dot{\mathcal{G}}\bigg(\frac{R}{\mathcal{G}}\bigg)^{2}\bigg].
\end{eqnarray}
Here, the Euler-Lagrange equations become
\begin{eqnarray}\nonumber &&
m(2m^2+3m-2)A^{2m-2}\dot{A}^{2}\frac{R}{\mathcal{G}}+2m(m+2)A^{2m-1}\ddot{A}\frac{R}{\mathcal{G}}+2m(m+2)A^{2m-1}\dot{A}\frac{\dot{R}}{\mathcal{G}}\\\nonumber
&&-2m(m+2)A^{2m-1}\dot{A}\frac{R\dot{\mathcal{G}}}{\mathcal{G}^2}+(2m+1)A^{2m}\frac{\ddot{R}}{\mathcal{G}}
-2(2m+1)A^{2m}\frac{\dot{R}\dot{\mathcal{G}}}{\mathcal{G}^2}-2(m+1)A^{2m}\frac{R\ddot{\mathcal{G}}}{\mathcal{G}^2}
\\\nonumber
&&+2(2m+1)A^{2m}\frac{R\dot{\mathcal{G}^2}}{\mathcal{G}^3}
+16m^2(m-1)A^{2m-3}\dot{A}^3\frac{R\dot{R}}{\mathcal{G}^2}+24m^2A^{2m-2}\dot{A}\ddot{A}\frac{R\dot{R}}{\mathcal{G}^2}
\\\nonumber&&+12m^2(m-1)A^{2m-3}\dot{A}^2\frac{\dot{R}^2}{\mathcal{G}^2}+12m^2A^{2m-2}\dot{A}^2\frac{R\ddot{R}}{\mathcal{G}^2}
-48m^2A^{2m-2}\dot{A}^2\frac{R\dot{R}\dot{\mathcal{G}}}{\mathcal{G}^3}
\\\nonumber&&-12m^2(m-1)A^{2m-3}\dot{A}^3\frac{R^2\dot{\mathcal{G}}}{\mathcal{G}^3}-24m^2A^{2m-2}\dot{A}\ddot{A}\frac{R^2\dot{\mathcal{G}}}{\mathcal{G}^3}
-12m^2A^{2m-2}\dot{A}^2\frac{R^2\ddot{\mathcal{G}}}{\mathcal{G}^3}\\&&+36m^2A^{2m-2}\dot{A}^2\frac{R^2\dot{\mathcal{G}^2}}{\mathcal{G}^4}=0,
\end{eqnarray}
\begin{eqnarray}
(2m+1)A^{2m}\ddot{A}+3m^2A^{2m-1}\dot{A}^2+8m^2(m-1)A^{2m-3}\dot{A}^4\frac{R}{\mathcal{G}}+12m^2A^{2m-2}\dot{A}^2\ddot{A}\frac{R}{\mathcal{G}}=0,
\end{eqnarray}
and the energy condition (\ref{energycondition}) takes the form
\begin{eqnarray}\nonumber &&
m(m+2)A^{2m-1}\dot{A}^2R+(2m+1)A^{2m}\dot{A}\dot{R}-(2m+1)A^{2m}\dot{A}\frac{R\dot{\mathcal{G}}}{\mathcal{G}}
+12m^2A^{2m-2}\dot{A}^3\frac{R\dot{R}}{\mathcal{G}}\\&&-12m^2A^{2m-2}\dot{A}^3\frac{R^2\dot{\mathcal{G}}}{\mathcal{G}^2}=0.\label{energyequation}
\end{eqnarray}
By putting the corresponding values of $R$ and $\mathcal{G}$ and
using Eq.(\ref{energyequation}), we get
\begin{eqnarray}\nonumber&&
12m^4(m^4-5m^3+2m+2)\dot{A}^8+m(132m^5-52m^4-41m^3-57m^2\\\nonumber&&-53m-10)A^2\dot{A}^4\ddot{A}^3
+(16m^4+8m^3-12m^2-10m-2)A^4\ddot{A}^4+\\\nonumber&&(80m^5+36m^4+40m^3+61m^2+24m+2)A^3\dot{A}^2\ddot{A}^3+m(4m^4+8m^3
\\\nonumber&&-3m^2-7m-2)A^3\dot{A}^3\ddot{A}\dddot{A}+m^2(92m^5-110m^4-45m^3-44m^2+14m\\&&+12)A\dot{A}^6\ddot{A}+m^2(10m^2+29m^3+24m^2+14m+4)A^2\dot{A}^5\dddot{A}=0.\label{const}
\end{eqnarray}

\subsection{Exponential law solutions}

Here we assume the metric coefficients in exponential law form, i.e.,
\begin{equation}\label{123}
A=e^{\varphi t},
\end{equation}
where $\varphi$ is an arbitrary constant. The solution metric for this case is
\begin{equation}
ds^2=dt^2-e^{2\varphi t}dx^2-e^{2m\varphi t}(dy^2+dz^2).
\end{equation}Using  Eq.(\ref{123}) in Eq.(\ref{const}), we get the constraint equation
\begin{align}(12m^8+32m^7+32m^6+40m^5+23m^4+16m^3+5m^2+2m)=0.\end{align}
The real solutions for Eq.(\ref{energyequation}) are
\begin{equation}m=0,~~m=-2.\end{equation}
Hence we obtain two cosmological solutions by considering $n=2$ in the $f(R,\mathcal{G})$ model. Many other solutions can be reconstructed by choosing some other values of the parameter. It is anticipated that a $\Lambda CDM$ universe may be generated for $m=1$ and some suitable value of $n$.

\subsection{Power law solutions}

We assume $A=t^\zeta,$ to extract a power law solution, where $\zeta$ is any non-zero real number. In this case, we obtain the constraint equation as
\begin{eqnarray}\nonumber
&&12m^4(m^4-5m^3+2m+2){\zeta}^4+2(m-1)(2m+1)^3({\zeta}^4-4{\zeta}^3+6{\zeta}^2-4{\zeta}+1)
\\\nonumber&&+m(2m+1)(-10+m(-33+m(9+m(-59+66m))))({\zeta}^4-2{\zeta}^3+{\zeta}^2)
\\\nonumber&&+m^2(12+m(14+m(-44+m(-45+2m(-55+46m)))))({\zeta}^4-{\zeta}^3)\\\nonumber&&+(2m+1)^2(2+m(16+m(-11+20m)))({\zeta}^4-3{\zeta}^3+3{\zeta}^2-{\zeta})\\\nonumber&&
+m^2(2+m)(2m+1)(2+m(2+5m))({\zeta}^4-3{\zeta}^3+2{\zeta}^2)\\&&+m(2m+1)^2(m^2+m-2)({\zeta}^4-4{\zeta}^3+5{\zeta}^2-2{\zeta})=0,\label{constraintequation}
\end{eqnarray}
In order to get a particular solution, we put ${\zeta}=-\frac{1}{3}$, so that
(\ref{constraintequation}) gives
\begin{equation}m=-2,~~m=-1,~~m=-\frac{2}{3}.\end{equation}
It is interesting to notice that for
$m=-2$, the solutions metric turns out to be
\begin{equation}
ds^2=dt^2-t^{\frac{-2}{3}}dx^2-t^{\frac{4}{3}}(dy^2+dz^2),
\end{equation}
which is same as the well-known Kasner's metric \cite{cata}.

\section{Final Remarks}

Specifically, we study $f(R,\mathcal{G})$ gravity, where the function $f$ consists of the Ricci scalar $R$ and
of the Gauss-Bonnet invariant $\mathcal{G}$.
There are many interesting aspects of Gauss-Bonnet theory which make it interesting for the researchers.
Gauss-Bonnet term is a specific combination of
curvature invariants that includes Ricci scalar, Ricci and Riemann tensors.
In fact, Gauss-Bonnet invariant naturally arises in the process of
quantum field theory regularization and renormalization of
curved spacetime. In particular, including $\mathcal{G}$ and $R$ in a bivariate function provides
a double inflationary scenario where the two acceleration phases are led by $\mathcal{G}$ and $R$ respectively \cite{mar}.
Moreover, the involvement of Gauss-Bonnet invariant may play an important role in the early time
expansion of universe as it is connected with the string theory and the trace anomaly \cite{cap.lau}.

In this paper, the Noether symmetry approach has been considered for an anisotropic
cosmological model in modified Gauss-Bonnet gravity.
The Lagrange multiplier approach allows us to deal with this
difficulties related to $f(R,\mathcal{G})$ model and reduces the Lagrangian into a
canonical form. In particular, we consider an LRS BI model and due to highly
non-linear and complicated field equations, we use a physical assumption $B=A^m$.
A system of PDEs has been constructed using Noether symmetries.
A detailed analysis of the determining equations is presented.
Firstly, a model $f(R,\mathcal{G})=f_0R^l+f_1\mathcal{G}^n$ is proposed and the corresponding Noether symmetries are investigated.
The results in this regard agree with \cite{cap} for special case when $m=1$. More importantly, we  have also recovered the Noether Symmetries for $f(R)$
 and $f(\mathcal{G})$ theories of gravity and the results agree with \cite{hus, shm12} for $f(R)$ gravity.

The last part of the paper deals with the reconstruction of some important cosmological solutions.  Exponential and power law solutions are reported for a well-known  $f(R,\mathcal{G})$ model, i.e., $f(R,\mathcal{G})=f_0R^n\mathcal{G}^{1-n}$ \cite{cap}. Especially, the Kasner's solution is recovered and it is anticipated that the familiar de-Sitter spacetime giving $\Lambda CDM$ cosmology
may be reconstructed for $m=1$ and some suitable value of $n$. In
this paper, we have just discussed a few examples where some
important Noether symmetries are reported. Many other cases can be
explored giving important cosmological solutions. It is worth
mentioning that our results agree with \cite{cap} for a special case
when $m=1$.

\end{document}